# A Mixed-Entropic Uncertainty Relation

Kamal Bhattacharyya* and Karabi Halder
Department of Chemistry, University of Calcutta, Kolkata 700 009, India

Abstract

We highlight the advantages of using simultaneously the Shannon and Fisher information measures in providing a useful form of the uncertainty relation for the position-momentum case. It does not require any Fourier transformation. The sensitivity is also noteworthy.

*Corresponding author (pchemkb@yahoo.com)

## 1. Introduction

Heisenberg's uncertainty relation (UR) involving standard deviations of position $x$ and reduced momentum $k$ ($k = p/\hbar$) is expressed as [1]

$$\sigma_x \sigma_k \geq \tfrac{1}{2}. \tag{1}$$

The relation has later been extended to two arbitrary non-commuting observables [2]. A recent survey [3] has paid considerable attention to several points of this UR. However, it was also realized that $\sigma_x$ is not always [4] a neat and physical measure of 'uncertainty' associated with the mean value of $x$. Such situations also include probability distributions (PD) like a Lorentzian for which $\sigma_x = \infty$ and distributions with multiple peaks. To combat the latter, an 'equivalent width' concept [5] was put forward long back, but it did not gain much popularity. Instead, another form of the UR by involving Shannon entropy has received sufficient curiosity. This UR is succinctly represented as [6 – 9]

$$S_S(x) + S_S(k) \geq 1 + \ln(\pi), \tag{2}$$

where $S_S(x)$ refers to the Shannon entropy in position space and is defined by

$$S_S(x) = -\langle \ln(P(x)) \rangle, \tag{3}$$

with $P(x)$ as the normalized PD. A similar definition for $S_S(k)$ follows with the corresponding PD $\bar{P}(k)$ in $k$-space. If $P(x) = |\Psi(x)|^2$, then $\bar{P}(k) = |\Phi(k)|^2$ such that functions $\Psi(x)$ and $\Phi(k)$ are Fourier transforms (FT) of each other:

$$\begin{aligned}\Psi(x) &= (1/\sqrt{2\pi}) \int_{-\infty}^{\infty} \Phi(k) \exp[ikx]\, dk; \\ \Phi(k) &= (1/\sqrt{2\pi}) \int_{-\infty}^{\infty} \Psi(x) \exp[-ikx]\, dx.\end{aligned} \tag{4}$$

The entropic UR (EUR), relation (2), has also been extended to general non-commuting observables [10].

The Shannon entropy has received considerable popularity in a wide variety of contexts (see, e.g., [11 – 16] and references quoted therein) starting from statistical mechanics [11, 12] to polymer chemistry [13], thermodynamics [14], quantum mechanics [15] and quantum chemistry [16]. So, EUR in the form (2) has received wide acceptance [4, 17]. As the next step, therefore, it has become natural to look for similar relations with other information measures. Thus, Rényi entropy has received some attention [18 – 20]; the relevance of Tsallis entropy in the context of EUR has been explored [21 – 23];

finally, role of the Fisher information measure has also been noted [24, 25]. Still, one observes that it is EUR (2) that is the most popular one.

The main difficulty with (2), however, is the FT part, unlike (1). More often than not, the transform is obtained only after cumbersome exercises. So, we like to modify (2) in such a way that any FT is avoided. We achieve this by recalling $I_F(x)$, the Fisher information in position space, a function of which replaces the $S_S(k)$ term in (2). This makes a concomitant change at the right side of (2) as well. But, we shall see that, while a Gaussian yields the equality here too, our form is sometimes more sensitive than (2).

**2. The relation**

The Fisher information measure is defined by

$$I_F(x) = \langle (d \ln(P(x)/dx)^2 \rangle . \tag{5}$$

For a real quantum-mechanical wave function, it is easy to check [15] that

$$I_F(x) = 4\sigma_k^2 . \tag{6}$$

Coupled with (1), (6) yields the well-known Cramer-Rao inequality [24]. In this context, it is essential to point out that Hall [24] discussed at some length on the genesis of following two inequalities:

$$\sqrt{2\pi e} \; \sigma_x \geq \exp[S_S(x)]; \tag{7a}$$

$$\exp[S_S(x)] \geq \sqrt{2\pi e / I_F(x)} . \tag{7b}$$

Of these, the first one [(7a)] has received considerable concern because it shows directly that (2) offers a tighter inequality than (1). Indeed, this feature alone provides a strong motivation in switching over to (2) from (1). Need less to mention, the exact equality in either case follows for a Gaussian distribution.

We shall, on the other hand, concentrate on the second inequality 7(b) on which only scanty attention was paid. A slight rearrangement leads us to

$$S_S(x) + \tfrac{1}{2}\ln(I_F(x)) \geq \tfrac{1}{2}\ln(2\pi e) . \tag{8}$$

If we compare (8) with (2), it becomes apparent that the second term at the left of (8) accounts, in effect, for $S_S(k)$. This identification has a nice physical appeal [15]. As the Fisher information in position space increases, the same in $k$-space decreases and hence the entropy increases. Effectively, thus, (8) becomes the new EUR. Here, the FT part is avoided naturally. One can put (8) more succinctly as

$$S_S(x) + \bar{S}_F(k) \geq \tfrac{1}{2}\ln(2\pi e) \tag{9}$$

where

$$S_S(k) \sim \bar{S}_F(k) = \tfrac{1}{2}\ln(I_F(x)). \tag{10}$$

Note that we have used here $\bar{S}_F(k)$ to denote a derived Fisher entropy. One might, for example, define a true Fisher entropy in $k$-space as

$$S_F(k) = -\ln(I_F(k)), \tag{11}$$

but that will again invite the FT, and hence is of little use here.

## 3. Some advantages

Let us choose a few test cases to see how quickly one can compute the desired quantities:

**Case (i):** For $P(x) = (2/L)\sin^2(n\pi x/L)$ with $x$ in $(0, L)$, we find

$$S_S(x) + \bar{S}_F(k) \geq 2\ln(2) + \ln(\pi) - 1 + \ln(n) \approx 1.53 + \ln(n). \tag{12}$$

This PD refers to the quantum-mechanical particle in a 1-d box model. Here, we notice that the entropy sum increases linearly with the logarithm of the quantum number. At large $n$, this is virtually equal to the number of zeroes of the PD. Indeed, we may appreciate that, while (12) is exact, such a dependence of the left side on the logarithm of the number of zeroes is a general semiclassical result for bound stationary states of any 1-dimensional potential. Actually, on employing the Wilson-Sommerfeld quantization rule, one finds [15] that

$$\bar{S}_S(x) + \bar{S}_F(k) = \ln(2\pi) + \ln(n), \tag{13}$$

where

$$\bar{S}_S(x) = -\ln(\langle P(x) \rangle) \tag{14}$$

and it satisfies (see also Sec. IV of ref. [15])

$$S_S(x) \geq \bar{S}_S(x). \tag{15}$$

We thus note two more aspects of the problem. First, the appearance of $\bar{S}_F(k)$ in the present context is a natural one. Secondly, (13) and (15) yield

$$S_S(x) + \bar{S}_F(k) \geq \ln(2\pi) \tag{16}$$

for the lowest ($n = 1$) quantum state. While semiclassical results are usually valid in the $n \to \infty$ limit, (16) is not too far from the exact inequality (8). In view of the current interest in $I_F$ for Schrödinger energy eigenfunctions [26], such a result is worth mentioning.

**Case (ii):** With $P(x) = (\alpha/\pi)(\alpha^2 + x^2)^{-1}$ and $x$ in (-∞, ∞), it turns out that

$$S_S(x) + \overline{S}_F(k) \geq \tfrac{3}{2}\ln(2) + \ln(\pi) \approx 2.18. \tag{17}$$

We talked about this Lorentzian PD earlier where the UR (1) fails to perform. Here, however, we notice how quickly one obtains a neat result.

**Case (iii):** We next choose $P(x) = 4\pi(\alpha/\pi)^{3/2} x^2 \exp[-\alpha x^2]$ with $x$ in (0, ∞). One finds here that

$$S_S(x) + \overline{S}_F(k) \geq \tfrac{1}{2}\ln(6\pi) + \gamma - \tfrac{1}{2} \approx 1.55, \tag{18}$$

where $\gamma$ is the Euler constant. This PD corresponds to the classical Maxwell speed distribution and, therefore, (18) shows the applicability of EUR to such a classical case too. Indeed, to incorporate such classical PD, we have deliberately avoided any reference to $\hbar$ at the outset and used the variable $k$ instead of $p$.

**Case (iv):** We finally choose $P(x) = (2\alpha^3/\pi)(\alpha^2 + x^2)^{-2}$ with $x$ in (-∞, ∞). This PD leads one to the result

$$S_S(x) + \overline{S}_F(k) \geq \tfrac{7}{2}\ln(2) + \ln(\pi) - 2 \approx 1.57. \tag{19}$$

This particular example has a special appeal. Going over to $\overline{P}(k)$, we obtain for this case

$$S_S(x) + S_S(k) = 3\ln(2) + \ln(\pi) - 1 \approx 1.079 + \ln(\pi). \tag{20}$$

This result may be contrasted with (19). One notes that this non-Gaussian PD reveals a deviation of about 10.7% in the EUR (8) from the result for a Gaussian while (20) shows that the value departs from the equality only by 3.7% when one opts for EUR (2). Such an outcome brings to light that, in situations, form (8) may be more sensitive than (2) as well. Actually, this deviation should have a direct link with the overlap of a chosen function with a Gaussian. Further work along this direction may be useful.

**4. Conclusion**

In summary, the use of mixed entropies in constructing EUR has not been tried so far. We have found here a form of EUR [see (8)] by invoking simultaneously the Shannon and Fisher information measures. It is simple and applicable to both classical and quantum PD. It does not require any FT. Sometimes, it is more sensitive than the parent form (2). Its semiclassical relevance in the context of Schrödinger eigenvalue problems is also notable. Further exploration with this form vis-à-vis other forms of EUR may be worthwhile. The derived Fisher entropy in *k*-space may also find other interesting applications.